\documentclass[conference]{IEEEtran}
\IEEEoverridecommandlockouts
\usepackage{cite}
\usepackage{amsmath,amssymb,amsfonts}
\usepackage{algorithmic}
\usepackage{graphicx}
\usepackage{textcomp}
\usepackage{xcolor}
\usepackage[T1]{fontenc}
\usepackage{multirow}

\def\BibTeX{{\rm B\kern-.05em{\sc i\kern-.025em b}\kern-.08em
    T\kern-.1667em\lower.7ex\hbox{E}\kern-.125emX}}
\begin{document}

\title{Generalizable Blood Pressure Estimation from Multi-Wavelength PPG Using Curriculum-Adversarial Learning\\

}
\author{
\IEEEauthorblockN{
Zequan Liang$^{1}$,
Ruoyu Zhang$^{2}$,
Wei Shao$^{1}$,
Mahdi Pirayesh Shirazi Nejad$^{2}$,\\
Ehsan Kourkchi$^{2}$,
Setareh Rafatirad$^{1}$,
Houman Homayoun$^{2}$
}\\
\IEEEauthorblockA{
$^{1}$Department of Computer Science, University of California, Davis, Davis, CA, U.S.A.\\ 
$^{2}$Department of Electrical and Computer Engineering, University of California, Davis, Davis, CA, U.S.A.\\ 
\{zqliang, ryuzhang, wayshao, pirayesh, ekay, srafatirad, hhomayoun\}@ucdavis.edu}}





\maketitle

\begin{abstract}

Accurate and generalizable blood pressure (BP) estimation is vital for the early detection and management of cardiovascular diseases. In this study, we enforce subject-level data splitting on a public multi-wavelength photoplethysmography (PPG) dataset and propose a generalizable BP estimation framework based on curriculum-adversarial learning. Our approach combines curriculum learning, which transitions from hypertension classification to BP regression, with domain-adversarial training that confuses subject identity to encourage the learning of subject-invariant features. Experiments show that multi-channel fusion consistently outperforms single-channel models. On the four-wavelength PPG dataset, our method achieves strong performance under strict subject-level splitting, with mean absolute errors (MAE) of 14.2mmHg for systolic blood pressure (SBP) and 6.4mmHg for diastolic blood pressure (DBP). Additionally, ablation studies validate the effectiveness of both the curriculum and adversarial components. These results highlight the potential of leveraging complementary information in multi-wavelength PPG and curriculum-adversarial strategies for accurate and robust BP estimation.

\end{abstract}


\section{Introduction}

Blood pressure (BP) is a vital physiological parameter for assessing cardiovascular health. Accurate and continuous BP monitoring is essential for the early detection and management of hypertension, which is a prevalent health problem worldwide \cite{joint1997report}. Conventional non-invasive BP monitoring methods, such as cuff-based oscillometric or auscultatory techniques, are widely used in clinical settings. However, these methods are often uncomfortable and unsuitable for long-term or ambulatory use \cite{picone2017accuracy}. Recent advances in wearable technology have introduced photoplethysmography (PPG) sensors for continuous health monitoring. In particular, this non-invasive optical method captures blood volume changes, offering a compact and low-cost solution for cuff-less BP estimation \cite{ding2017pulse}. 

\begin{figure}[ht]
    \centering
    \includegraphics[width= \linewidth,height=0.13\textheight]{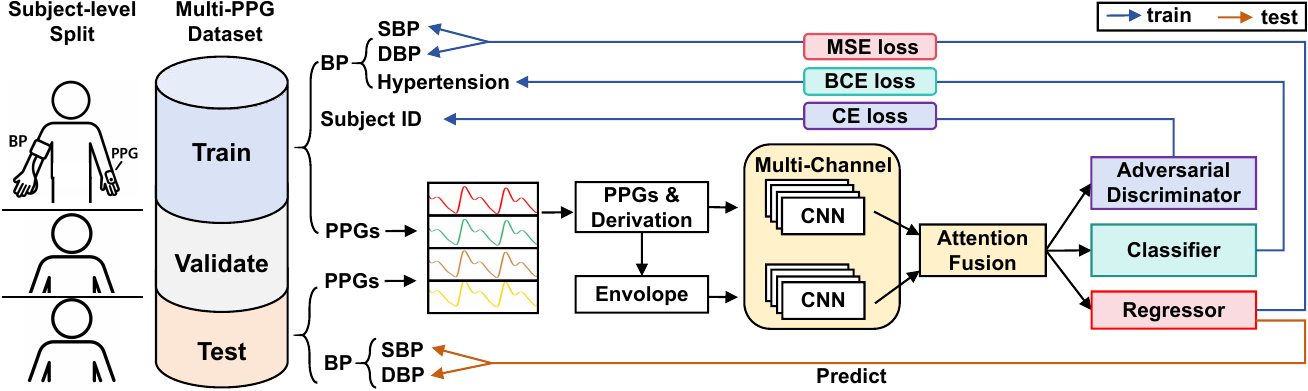}
    \caption{Overview of blood pressure estimation from multi-wavelength PPG}
    \label{fig:big_picture}
\end{figure}

Motivated by these advantages, researchers have increasingly explored machine learning approaches for BP estimation from PPG signals. However, there are two major limitations. First, most existing studies rely solely on single-wavelength PPG, neglecting the additional information from multi-wavelength signals \cite{esmaelpoor2020multistage}. Second, most studies use improper data splitting strategies, such as random segment-level splitting, leading to data leakage and overestimated performance due to overlap between training and test subjects \cite{yoshizawa2023investigation}.

To address these challenges, we propose a generalizable BP estimation framework using multi-wavelength PPG signals. Our main contributions are as follows: 1) We utilize a public dataset \cite{cui2024acnn} containing four-wavelength PPG channels and BP measurements to extract complementary features across different wavelengths. 2) To eliminate leakage, we enforce subject-level data splitting and introduce domain-adversarial training of Neural Networks (DANN) \cite{ganin2016domain} to learn subject-invariant features, thereby enhancing generalization for BP estimation. 3) We adopt a curriculum learning \cite{graves2017automated}, where the model is first trained to classify hypertension status before gradually transitioning to BP regression, guiding it to capture coarse-level clinical patterns before making fine-grained predictions. We refer to this combined approach as Curriculum-Adversarial Learning.

\section{Methodology} 

\subsection{Dataset}
\begin{figure}[t]
    \centering
    \includegraphics[width= \linewidth]{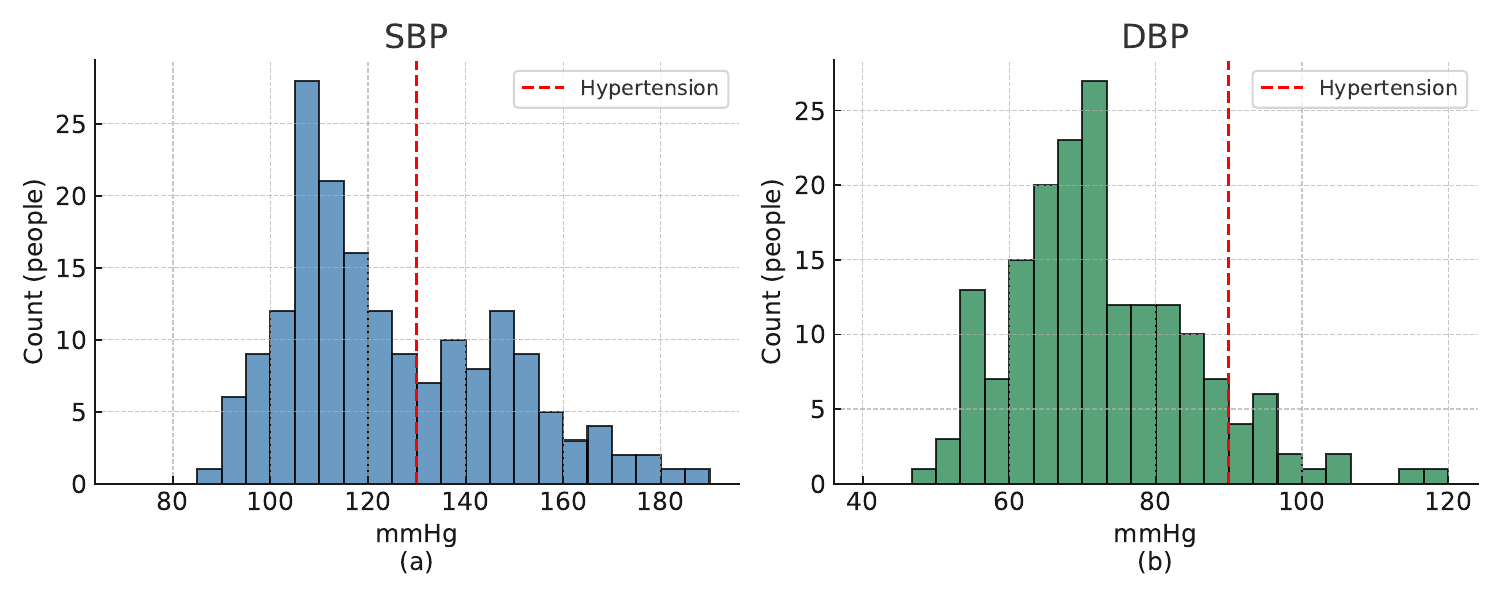}
    \caption{Blood pressure distribution (a) SBP (b) DBP}
    \label{fig:BP_distribution}
\end{figure}

Fig.\ref{fig:big_picture} illustrates the overall pipeline of our proposed BP estimation framework based on multi-wavelength PPG signals (660nm, 730nm, 850nm, and 940nm). Our approach is built upon the public dataset introduced by Cui et al.~\cite{cui2024acnn}, which provides one-minute four-channel fingertip PPG recordings, followed immediately by two reference blood pressure measurements of systolic (SBP) and diastolic (DBP) values. The dataset includes 180 subjects, 66 of whom are diagnosed with hypertension (SBP $\geq$ 130 mmHg or DBP $\geq$ 90 mmHg). The distribution of BP is presented in Fig~\ref{fig:BP_distribution}.

However, the original study \cite{cui2024acnn} proposing this public dataset and the ACNN-BiLSTM model does not report its data splitting strategy, raising concerns of potential data leakage due to shared subject-specific information between training and testing sets. To address this, we adopt strict subject-level data partitioning, dividing the dataset into training, validation, and testing sets with no subject overlap, ensuring a fair and generalizable evaluation. 

\subsection{Data Preprocessing and Feature Extraction}

We adopt the deep-learning–based algorithm proposed by Kyung et al.~\cite{kyung2023deep}, which uses an attention mechanism to leverage multi-wavelength PPG inputs. This method was originally developed on a private dataset that is not publicly available. We first apply their data preprocessing and feature extraction pipeline, excluding the finger-pressure branch, and then integrate our curriculum-adversarial learning strategy.

\begin{figure}[t]
    \centering
    \includegraphics[width= \linewidth]{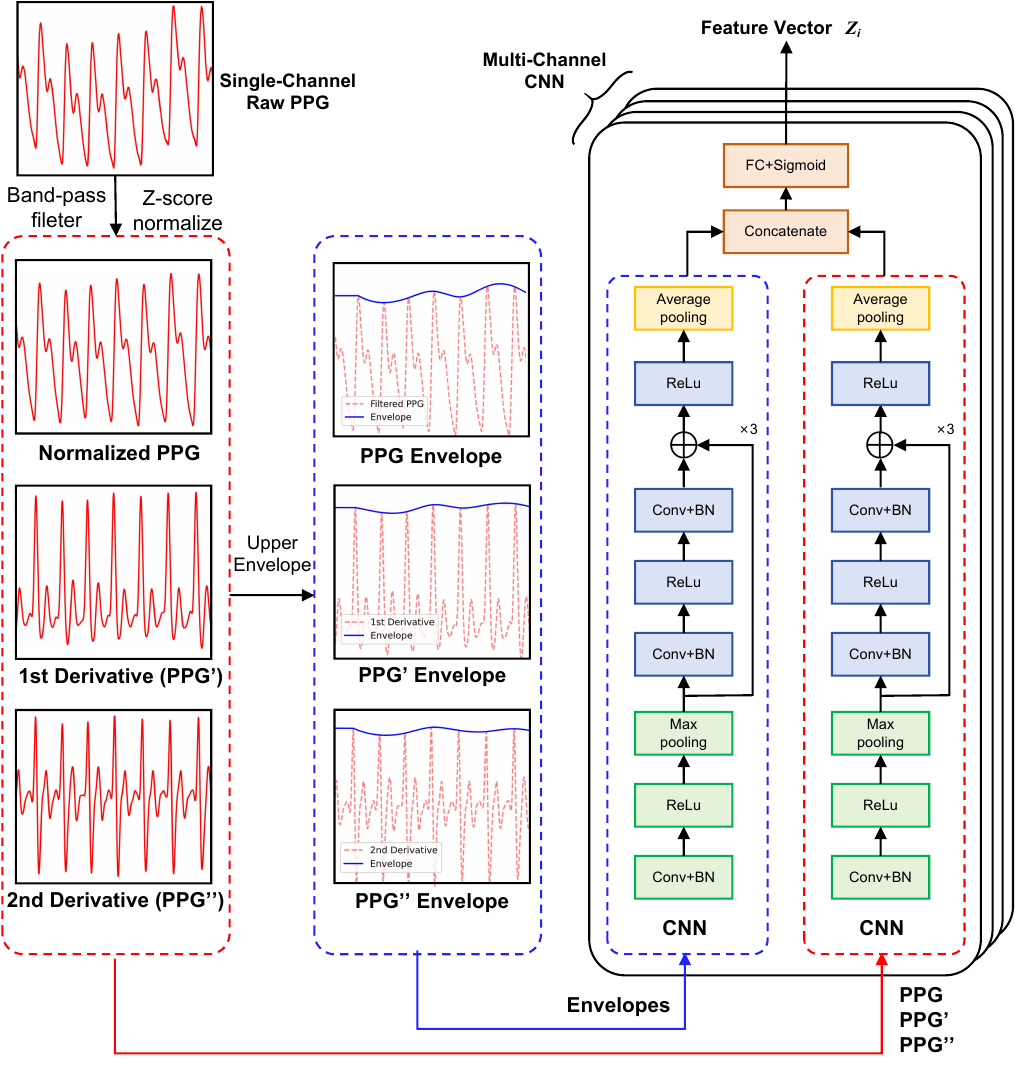}
    \caption{Data preprocessing and multi-channel feature extraction}
    \label{fig:BP_process}
\end{figure}

As shown on the left side of Fig.~\ref{fig:BP_process}, we preprocess each PPG signal by applying a band-pass filter within 0.5–8 Hz and Z-score normalization. We then compute the first and second-order derivatives to capture temporal dynamics. Additionally, for each original PPG signal and its derivatives, we extract the upper envelope to capture peaks.

Each PPG channel is processed through two parallel convolutional neural networks (CNN): one that takes the normalized PPG and its derivatives, and another that takes the corresponding envelopes. This forms a total of six input streams per channel. As shown on the right side of Fig.~\ref{fig:BP_process}, each CNN consists of multiple layers of convolution (Conv), batch normalization (BN), ReLU activation, and pooling. These networks extract informative temporal features independently from each wavelength of PPG, producing a feature vector $Z_i$ for the $i$-th PPG channel through a fully connected (FC) layer.

\subsection{Multi-Channel Fusion}
As shown in Fig.~\ref{fig:BP_model}, each feature vector $Z_i$ is extracted from its corresponding CNN branch. These vectors are then passed through an attention layer that assigns a soft weight $w_i$ to each channel, and the fused feature is obtained as their weighted sum. It serves as the input to three separate heads:



\begin{figure}[t]
    \centering
    \includegraphics[width= 0.9\linewidth]{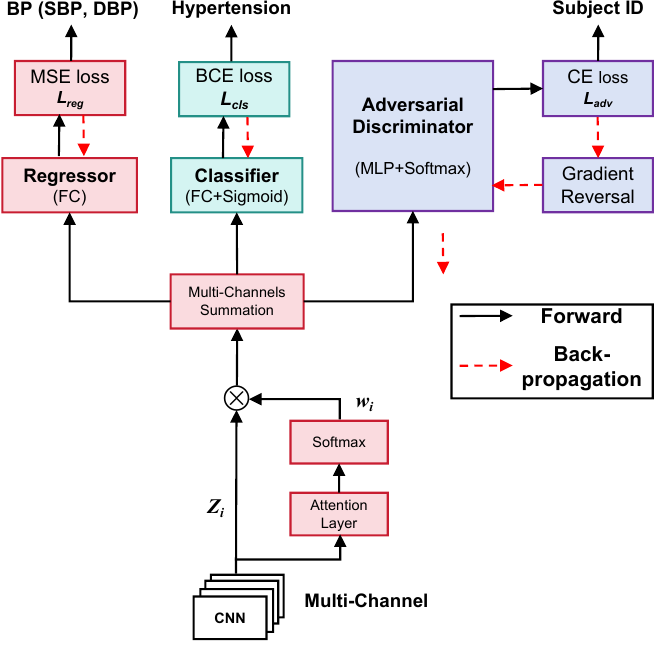}
    \caption{Multi-channel fusion and curriculum-adversarial learning}
    \label{fig:BP_model}
\end{figure}

\begin{itemize}

\item A regressor that estimates SBP and DBP values.

\item A binary classifier that predicts whether the subject is hypertensive (SBP $\geq$ 130 mmHg or DBP $\geq$ 90 mmHg) as part of the curriculum learning.

\item An adversarial discriminator that identifies the subject identity.

\end{itemize}

\subsection{Domain-Adversarial Training}

Inspired by unsupervised domain adaptation \cite{liu2025modfinity}, we treat different subjects as distinct domains. To further enhance generalization across subjects, we incorporate a domain-adversarial discriminator \cite{ganin2016domain} that attempts to classify subject identities. A gradient reversal layer (GRL) is employed during backpropagation to invert gradients, thereby encouraging the preceding multi-channel CNNs to learn subject-invariant features that confuse the discriminator. The adversarial loss is computed using categorical cross-entropy (CE):

\begin{equation}
L_{\text{adv}} = CE(\hat{y}_{\text{sbj}}, y_{\text{sbj}})
\end{equation}
where $\hat{y}_{\text{sbj}}$ is the predicted subject identity from the discriminator, and $y{_\text{sbj}}$ is the true subject identity. 


\subsection{Curriculum Learning}
Additionally, we adopt a curriculum learning strategy to guide the model from coarse-level classification to precise BP estimation. In the first stage, we train the binary classifier to distinguish hypertensive ($c$ = 1) and normotensive ($c$ = 0) subjects using the binary cross-entropy (BCE) loss:

\begin{equation}
L_{\text{cls}} = -[c \cdot \log(\hat{c}) + (1 - c) \cdot \log(1 - \hat{c})]
\end{equation}
where $\hat{c}$ denotes the predicted probability of hypertension from the classifier.

Once the classification loss stabilizes, we gradually shift toward a regression task that estimates SBP and DBP values using the mean squared error (MSE) loss:

\begin{equation}
L_{\text{reg}} = \frac{1}{2} \left[(\hat{y}_{\text{sbp}} - y_{\text{sbp}})^2 + (\hat{y}_{\text{dbp}} - y_{\text{dbp}})^2\right]
\end{equation}
where $\hat{y}_\text{sbp}$ and $\hat{y}_\text{dbp}$ are the predicted SBP and DBP values from the regressor. $y_{\text{sbp}}$ and $y_{\text{dbp}}$ are the true values.

The total loss function integrates the three objectives:

\begin{equation}
L = \lambda_1 L_{\text{reg}} + (1 - \lambda_1) L_{\text{cls}} - \lambda_2 L_{\text{adv}}
\end{equation}

Here, $\lambda_1$ is an adaptive weight defined as the ratio of the current epoch to the total number of training epochs, gradually shifting the learning focus from classification to regression. $\lambda_2$ is a fixed weight that controls the contribution of the adversarial loss. 

These objectives are jointly optimized by using this total loss function to train the overall model. During validation and testing, only the regression head is retained to estimate SBP and DBP values. 

\section{Experiment Results}

\subsection{Experiment Setting}
We evaluate our method on the public multi-wavelength PPG and BP dataset \cite{cui2024acnn}. For each subject, we extract the middle 30 seconds of PPG data. Subjects are split at the subject level into training and test sets using a 4:1 ratio to ensure no overlap across splits. Additionally, we perform 5-fold cross-validation within the training set. All models are implemented in PyTorch and trained using the Adam optimizer with a learning rate of 0.001 and a batch size of 32. Each model is trained for 100 epochs. We initialize the loss weights as $\lambda_1$ = 0 and $\lambda_2$ = 1, where $\lambda_1$ is gradually increased with the number of training epochs.

We evaluate BP estimation performance using four standard metrics. MAE (Mean Absolute Error) measures the average magnitude of prediction errors in mmHg, where lower values indicate better accuracy. BHS-5, BHS-10 and BHS-15 denote the percentage of samples with absolute errors below 5 mmHg, 10 mmHg and 15 mmHg, respectively, where higher values indicate better performance. They follow the British Hypertension Society (BHS) grading protocol \cite{williams2004british}.

\subsection{Effect of Multi-Wavelength Fusion}
To evaluate the effectiveness of leveraging multi-wavelength PPG signals, we compare the performance of our curriculum-adversarial learning–based multi-channel CNN fusion model across individual channels and the full multi-channel fusion. 

As shown in Table~\ref{tab:bp_channels}, the best-performing single channel for SBP estimation is 660nm, while the 940nm channel performs best for DBP. However, the all-channel model achieves superior results, yielding the lowest MAE as well as the highest BHS-5 and BHS-10 scores. These results demonstrate that integrating multi-wavelength PPG signals enhances BP prediction performance by leveraging complementary PPG features.



\begin{table}[t]
\centering
\caption{BP Predictions of Our Approach on Different PPG Channels}
\label{tab:bp_channels}
\setlength{\tabcolsep}{5.5pt}
\begin{tabular}{|c|c|c|c|c|c|c|c|c|}
\hline
\multirow{3}{*}{\textbf{Channel}} 
& \multicolumn{2}{c|}{\textbf{MAE}} 
& \multicolumn{2}{c|}{\textbf{BHS-5}} 
& \multicolumn{2}{c|}{\textbf{BHS-10}} 
& \multicolumn{2}{c|}{\textbf{BHS-15}} \\
& \multicolumn{2}{c|}{\textbf{(mmHg)}} 
& \multicolumn{2}{c|}{\textbf{(\%)}}
& \multicolumn{2}{c|}{\textbf{(\%)}}
& \multicolumn{2}{c|}{\textbf{(\%)}} \\
\cline{2-9}
& \textbf{\textit{SBP}} & \textbf{\textit{DBP}} 
& \textbf{\textit{SBP}} & \textbf{\textit{DBP}} 
& \textbf{\textit{SBP}} & \textbf{\textit{DBP}} 
& \textbf{\textit{SBP}} & \textbf{\textit{DBP}} \\
\hline
660\,nm & 14.9 & 8.6 & \textbf{24.2} & 48.5 & 39.4 & 69.7 & 54.5 & 75.8 \\
\hline
730\,nm & 15.5 & 8.2 & 21.2 & 21.2 & 36.4 & 66.7 & 48.5 & 90.9 \\
\hline
850\,nm & 15.3 & 7.7 & 21.2 & 36.4 & 45.5 & 72.7 & 54.5 & 81.8 \\
\hline
940\,nm & 16.2 & 7.3 & 21.2 & 36.4 & 36.4 & 69.7 & 54.5 & \textbf{93.9} \\
\hline
All     & \textbf{14.2} & \textbf{6.4} & \textbf{24.2} & \textbf{51.5} & \textbf{51.5} & \textbf{75.8} & \textbf{57.6} & 90.9 \\
\hline
\end{tabular}
\end{table}

\subsection{Comparison with Existing Methods on Multi-Wavelength PPG}

To validate the effectiveness of our proposed method, we compare it against several existing machine learning approaches that utilize multi-wavelength PPG signals for BP estimation:

\begin{table}[t]
\centering
\caption{BP Prediction Comparison of Different Methods Using Multi-Wavelength PPG}
\label{tab:bp_comparison}
\setlength{\tabcolsep}{5pt}
\begin{tabular}{|c|c|c|c|c|c|c|c|c|}
\hline
\multirow{3}{*}{\textbf{Method}} 
& \multicolumn{2}{c|}{\textbf{MAE}} 
& \multicolumn{2}{c|}{\textbf{BHS-5}} 
& \multicolumn{2}{c|}{\textbf{BHS-10}} 
& \multicolumn{2}{c|}{\textbf{BHS-15}} \\
& \multicolumn{2}{c|}{\textbf{(mmHg)}} 
& \multicolumn{2}{c|}{\textbf{(\%)}} 
& \multicolumn{2}{c|}{\textbf{(\%)}} 
& \multicolumn{2}{c|}{\textbf{(\%)}} \\
\cline{2-9}
& \textbf{\textit{SBP}} & \textbf{\textit{DBP}} 
& \textbf{\textit{SBP}} & \textbf{\textit{DBP}} 
& \textbf{\textit{SBP}} & \textbf{\textit{DBP}} 
& \textbf{\textit{SBP}} & \textbf{\textit{DBP}} \\
\hline
A-BiLSTM & 20.2 & 11.5 & 15.2 & 30.3 & 36.4 & 45.5 & 51.5 & 69.7 \\
\hline
MLP         & 19.3 & 10.4 & 12.1 & 30.3 & 33.3 & 48.5 & 42.4 & 78.8 \\
\hline
CNN1D       & 17.9 &  7.5 & 15.2 & \textbf{51.5} & 36.4 & 66.7 & 51.5 & 81.8 \\
\hline
Multi-CNN   & 16.1 &  7.2 & 15.2 & 48.5 & 36.4 & 63.6 & \textbf{57.6} & 87.9 \\
\hline
This Study  & \textbf{14.2} & \textbf{6.4} & \textbf{24.2} & \textbf{51.5} & \textbf{51.5} & \textbf{75.8} & \textbf{57.6} & \textbf{90.9} \\
\hline
\end{tabular}
\end{table}

\begin{table}[t]
\centering
\caption{Ablation Study of Our Components on BP Prediction}
\label{tab:ablation_bp}
\setlength{\tabcolsep}{5.2pt}
\begin{tabular}{|c|c|c|c|c|c|c|c|c|c|}
\hline
\multirow{3}{*}{\textbf{cls}} 
& \multirow{3}{*}{\textbf{adv}} 
& \multicolumn{2}{c|}{\textbf{MAE}} 
& \multicolumn{2}{c|}{\textbf{BHS-5}} 
& \multicolumn{2}{c|}{\textbf{BHS-10}} 
& \multicolumn{2}{c|}{\textbf{BHS-15}} \\
& & \multicolumn{2}{c|}{\textbf{(mmHg)}} 
& \multicolumn{2}{c|}{\textbf{(\%)}} 
& \multicolumn{2}{c|}{\textbf{(\%)}} 
& \multicolumn{2}{c|}{\textbf{(\%)}} \\
\cline{3-10}
& & \textbf{\textit{SBP}} & \textbf{\textit{DBP}} 
  & \textbf{\textit{SBP}} & \textbf{\textit{DBP}} 
  & \textbf{\textit{SBP}} & \textbf{\textit{DBP}} 
  & \textbf{\textit{SBP}} & \textbf{\textit{DBP}} \\
\hline
             &              & 16.1 & 7.2 & 15.2 & 48.5 & 36.4 & 63.6 & 57.6 & 87.9 \\
\hline
             & \checkmark   & 17.1 & 7.1 &  9.1 & 48.5 & 30.3 & \textbf{78.8} & 51.5 & 87.9 \\
\hline
\checkmark   &              & 14.3 & 7.1 & \textbf{27.3} & 33.3 & 45.5 & 75.8 & \textbf{60.6} & \textbf{90.9} \\
\hline
\checkmark   & \checkmark   & \textbf{14.2} & \textbf{6.4} & 24.2 & \textbf{51.5} & \textbf{51.5} & 75.8 & 57.6 & \textbf{90.9} \\
\hline
\end{tabular}
\end{table}

\begin{itemize}

\item A-BiLSTM\cite{cui2024acnn}: Converts PPG signals into 2D continuous wavelet transform spectrograms and applies a hybrid attention-based CNN with bidirectional long short-term memory (BiLSTM) for BP regression. 

\item MLP\cite{botrugno2024ai}: Extracts hand-crafted features from PPG and applies a multilayer perceptron (MLP) for regression.

\item CNN1D\cite{botrugno2024ai}: Aggregates all PPG channels by square summation and applies a 1D CNN for BP estimation.

\item Multi-CNN\cite{kyung2023deep}: Applies multiple CNNs to various PPG channels and fuses the extracted features using attention. It serves as the backbone of our proposed model.

\item This study: Our proposed approach, which extends the Multi-CNN with curriculum-adversarial learning. 

\end{itemize}

As shown in Table~\ref{tab:bp_comparison}, our proposed approach outperforms all baselines in both SBP and DBP estimation. Specifically, it achieves the lowest MAE of 14.2mmHg for SBP and 6.4mmHg for DBP. It consistently achieves the best performance across all evaluation metrics. These results validate the effectiveness of our proposed curriculum-adversarial learning framework in leveraging multi-wavelength PPG signals for generalizable BP estimation.

\subsection{Ablation Study}

To evaluate the contribution of each proposed component, we conducted an ablation study by selectively removing the classifier (cls) and the adversarial discriminator (adv). 

Table~\ref{tab:ablation_bp} summarizes our ablation study, where $\checkmark$ denotes inclusion of the classification (cls) or adversarial (adv) component, and a blank cell indicates its removal. The full model incorporating both components achieves the lowest MAE for both SBP and DBP estimation, outperforming variants without either component. It also achieves the best result in at least one of SBP or DBP across all BHS metrics. These results further confirm that both the hypertension classification task and the adversarial discriminator play important roles.

\section{Discussion} 
With strict subject-level splitting and a uniform preprocessing pipeline, our proposed curriculum-adversarial learning strategy achieves better results compared to existing multi-wavelength PPG methods. The stricter protocol inevitably increases the reported error, but it provides a more realistic assessment of model performance. While the model does not consistently outperform all baselines across every evaluation metric, it shows potential for applications such as hypertension screening and long-term trend monitoring. Nevertheless, the accuracy still has room for improvement, especially under extreme blood pressure ranges and degraded signal quality.

\section{Conclusion and Future Work}
We proposed a generalizable blood pressure estimation framework using multi-wavelength PPG signals, enhanced by a curriculum-adversarial learning strategy. By combining coarse-to-fine curriculum learning with domain-adversarial training, our model learns both hypertension-level and subject-invariant features, while avoiding data leakage through strict subject-level splitting. Experimental results on a public four-wavelength PPG dataset demonstrate that our method effectively extracts richer physiological features from multi-wavelength signals and achieves improved accuracy and generalization in both SBP and DBP estimation. In future work, we plan to collect additional multi-wavelength PPG data and further enhance the model through signal quality assessment.

\bibliographystyle{ieeetr}
\bibliography{EMBC}

\end{document}